\documentclass[a4paper]{article}
\usepackage[dvipsnames]{xcolor}
\usepackage{INTERSPEECH2020}

\title{Phase-aware music super-resolution using generative adversarial networks}
\name{Shichao Hu$^{*}$, Bin Zhang$^{*}$, Beici Liang, Ethan Zhao, Simon Lui}
\address{
  Tencent Music Entertainment (TME), Shenzhen, 518057, China}
\email{\{shichaohu,robingzhang,beiciliang,ethanzhao,nomislui\}@tencent.com}
\begin{document}

\maketitle
%

\begin{abstract}
Audio super-resolution is a challenging task of recovering the missing high-resolution features from a low-resolution signal. To address this, generative adversarial networks (GAN) have been used to achieve promising results by training the mappings between magnitudes of the low and high-frequency components. However, phase information is not well-considered for waveform reconstruction in conventional methods. In this paper, we tackle the problem of music super-resolution and conduct a thorough investigation on the importance of phase for this task. We use GAN to predict the magnitudes of the high-frequency components. The corresponding phase information can be extracted using either a GAN-based waveform synthesis system or a modified Griffin-Lim algorithm. Experimental results show that phase information plays an important role in the improvement of the reconstructed music quality. Moreover, our proposed method significantly outperforms other state-of-the-art methods in terms of objective evaluations.
\end{abstract}

\section{Introduction}
Audio super-resolution (SR) deals with the problem of recovering high-resolution (HR) audio from a low-resolution (LR) input. This is also known as "bandwidth expansion". Traditionally, this technique is applied to telephone speech transmission where speech resolution is limited due to the low-bandwidth channels~\cite{nilsson2001avoiding}. There are also studies showing that artificially expanding the speech bandwidth improves the perception for hearing-impaired persons~\cite{liu2009effect}. In this paper, our study is mainly focused on SR tasks for music and aimed at improving the audio quality of LR music by predicting both the magnitudes and phase of the high frequency components (denoted as HFC). 

Conventional approaches for SR problems are based on signal processing methods, such as linear mapping from the LR to HR spectral characteristics using linear predictive coding (LPC) analysis~\cite{nakatoh2002generation}. The mapping method can be extended with more complex statistical techniques, for instance, Gaussian mixture models (GMMs)~\cite{park2000narrowband} and Hidden Markov Model (HMM)~\cite{jax2003artificial}.

With the development of deep learning, recent audio SR approaches based on deep neural networks (DNNs) present better performance than the conventional methods. Li et al.~\cite{li2015deep} used fully connected DNNs to learn the LR to HR mapping of spectral magnitude and then reconstruct the HR audio with the flipped phase of low frequency components (denoted as LFC). Kuleshov et al.~\cite{kuleshov2017audio} employed an auto-encoder network to generate the HR waveform from the LR waveform input. This time domain approach can avoid the phase problem and expand bandwidth through temporal interpolation. More recently, Eskimez et al.~\cite{eskimez2019speech,eskimez2019adversarial} introduced a novel method that estimates log-power spectrograms of the HFC using the generative adversarial network (GAN). This was inspired by the great success of GAN on generating highly realistic images~\cite{dong2015image, kim2016accurate}. Experimental results of such GAN method achieved the state-of-the-art performance in the frequency domain. 



To the best of our knowledge, existing literature using frequency-domain approach with DNNs has not fully considered the HFC phase to reconstruct HR signals. Considering the difficulty of phase modeling, directly mapping the HFC phase from LFC remains unsolved. Given the fact that phase plays an important role in speech perception, enhancement and synthesis~\cite{paliwal2011importance,mowlaee2016advances,ai2020neural}, we propose to use a recent MelGAN-based audio vocoder~\cite{kumar2019melgan} to predict the phase in Section~\ref{sec:proposed}. HR waveform can be generated using the predicted phase and the magnitude which is estimated using the GAN-based method mentioned above. Our proposed method outperforms the other state-of-the-art methods according to the evaluation results in Section~\ref{sec:results}. We further investigate the importance of phase in the music SR tasks through evaluating the quality of audio generated by different methods. Examples of the 
reconstructed music are available online \footnote{https://github.com/tencentmusic/TME-Audio-Super-Resolution-Samples}.
\let\thefootnote\relax\footnotetext{*Indicating Equal Contribution.}

\section{Related Works}
Recent studies have shown deep-learning-based methods are effective on audio SR tasks. This section reviews the recent advanced audio SR approaches and introduces phase estimation methods.

\subsection{Audio Super-Resolution with Deep Learning}
Audio SR with deep learning is usually approached from either frequency or time domain. One of the earliest frequency-domain approaches was proposed in \cite{li2015deep} using restricted Boltzmann machines as DNNs for training the mapping between magnitudes of LFC and HFC. A recent study introduced GANs for the purpose and shows the state-of-the-art results~\cite{eskimez2019speech,eskimez2019adversarial}. A typical time-domain approach was proposed recently ~\cite{kuleshov2017audio} which applied an auto-encoder network to generate HR waveform from the LR temporal signals.  

While the GANs-based spectra-mapping method shows promising results, the HFC phase for HR waveform reconstruction was not fully considered. In the existing works, it is artificially produced by flipping and repeating the LFC phase and adding a negative sign (denoted as FLIP). Since phase has been confirmed to effectively improve subjective perception on speech~\cite{paliwal2011importance,mowlaee2016advances}, we attempt to predict HFC magnitudes complemented by the estimated phase in order to recover the HR waveform with improvements in music SR tasks. 

\subsection{Phase Estimation}
\label{sec:phase_estimation}
Here two common methods for phase estimation are briefly introduced. One is Griffin-Lim algorithm (GLA), and the other one is based on speech vocoder. They are used and compared in our music SR tasks. 

GLA has been commonly used as a phase recovery method based on the consistency of spectrogram\cite{griffin1984signal,perraudin2013fast}. GLA method iteratively updates the complex-valued spectrogram while maintains the given magnitude ($\textbf{A}$). The process of GLA can be explained by the following equation (here we largely followed the notation used in \cite{masuyama2019deep}):
\begin{equation}
  \textbf{X}^{[m+1]} = P_{C}(P_A(\textbf{X}^{[m]})),
  \label{eq1}
\end{equation}
where $\textbf{X}^{[m]}$ denotes the complex-valued spectrogram updated through the $m$-th iteration. $P_{A}$ updates the phase and maintains the given magnitudes, i.e., $\textbf{A}$. $P_{C}$ calculates the corresponding consistent spectrograms. $P_{C}$ and $P_{A}$ are respectively given by:
\begin{equation}
  P_{C}(\textbf{X}) = \mathbb{F} \mathbb{F^{\dagger}}\textbf{X},
  \label{eq2}
\end{equation}
\begin{equation}
  P_{A}(\textbf{X}) = \textbf{A} \odot \textbf{X}\oslash |\textbf{X}|,
  \label{eq3}
\end{equation}
where $\mathbb{F}$ and $\mathbb{F^{\dagger}}$ are the forward short-time Fourier transform (STFT) and inverse STFT, respectively. $\odot$ is element-wise multiplication; $\oslash$ represents element-wise division and division by zero is replaced by zero. Through iterations, complex-valued spectrograms are consistently optimized by minimizing the difference between the energy of $\textbf{X}$ and $P_{C}(\textbf{X})$. To adapt GLA to our music SR tasks, we not only maintain the given LFC magnitude but also the LFC phase in Eq.~\ref{eq3}. Although GLA has been widely used for the past decades due to its simplicity, it requires many iterations and resulted reconstruction quality is not good enough. In the experiments of this study, the GLA is implemented with $100$ iterations, following the number of iterations used in \cite{masuyama2019deep}.

Recently, neural-network-based speech vocoder has been extensively developed for waveform synthesis over the past few years. {Although the neural-network-based vocoder has been generally developed for speech applications, it may have promising performance as well when applied to music audio modeling~\cite{Oord2016}. Here we decide to use a state-of-the-art method proposed in \cite{kumar2019melgan}, which generates audio waveform from mel-spectrogram using GANs (denoted as MelGAN). The MelGAN model consists of a fully convolutional generator network that generate raw waveform from mel-spectrogram input, and a discriminator network with a multi-scale architecture. 
In our task, a waveform can be generated using the mel-spectorgram concatenating the LFC of the input LR signal and the estimated HFC. The phase information of HFC can be therefore extracted from the generated audio.

\section{Proposed System}
\label{sec:proposed}

\subsection{Joint Framework}


We propose a joint music SR framework, mainly consisting of a ``magnitude component'' (MC) and a ``phase component'' (PC). Following the GANs architecture presented in \cite{eskimez2019adversarial}, MC predicts the magnitudes of HFC from the LFC. Magnitude prediction is done by a generator network with a convolutional auto-encoder architecture. Besides, a discriminator network is designed to distinguish the estimated magnitudes from real ones. To train MC, we used Wasserstein distance~\cite{arjovsky2017wasserstein} as GAN loss function to stabilize the training. As for PC, we used the the MelGAN model presented in \cite{kumar2019melgan} to convert mel-spectrogram into waveform, from which the phase of HFC can be estimated. 

\begin{figure}[t]
  \centering
  \includegraphics[width=\linewidth]{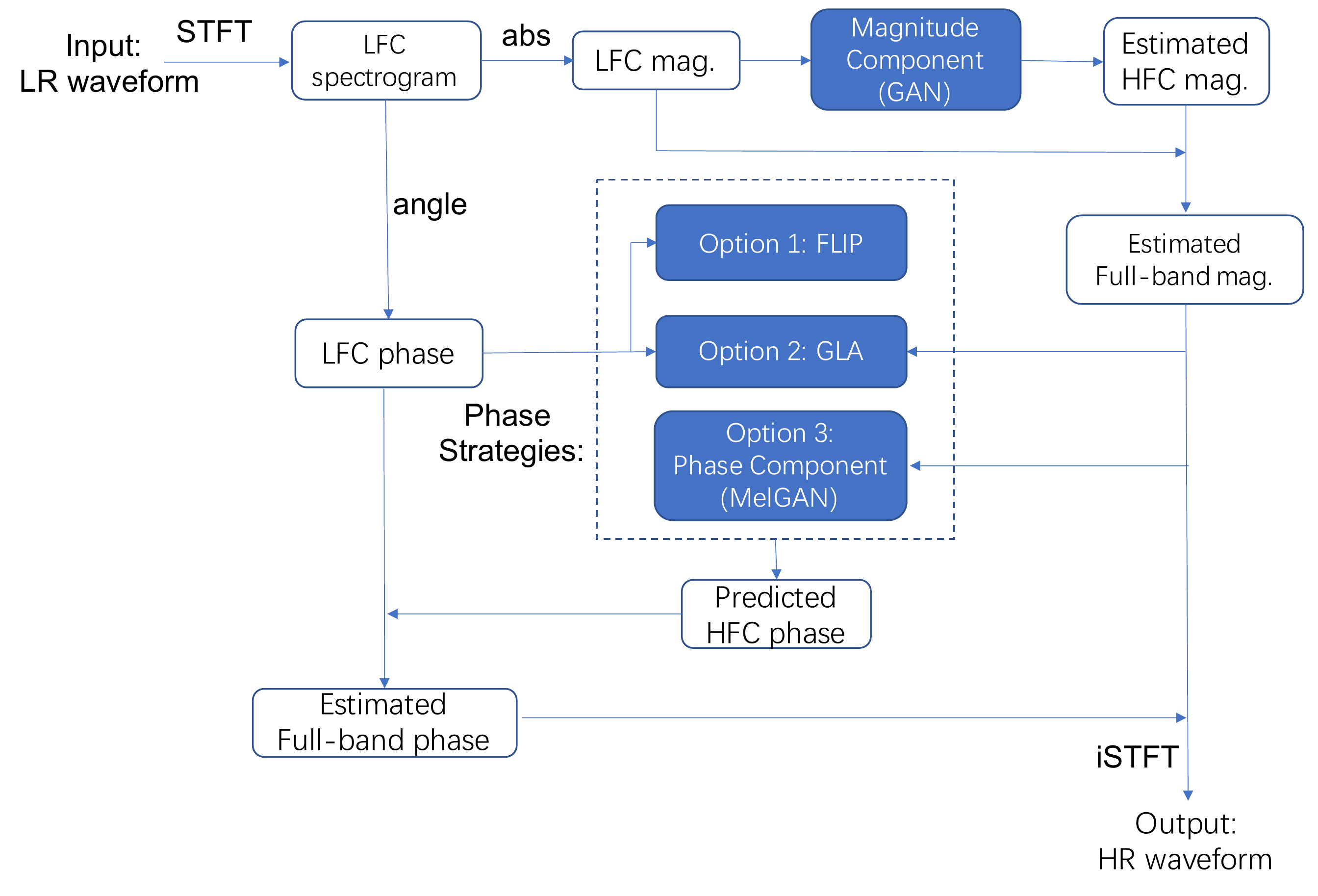}
  \caption{Schematic overview of the proposed system.}
  \label{fig:proposed-system}
\end{figure}
Fig.~\ref{fig:proposed-system} presents the whole joint framework. Both the magnitudes from MC and the phase from PC are used to obtain complex-valued spectrogram of high frequency bands. To estimate phase, PC with MelGAN is used in order to obtain a more accurate phase information for audio reconstruction. It is compared with another two strategies, namely FLIP and GLA. With this joint methods, quality of the reconstructed audio is expected to be improved. A detailed calculation process is illustrated as follows:

\begin{itemize}
\item Given the LR waveform input ($\textbf{x}_{LR}$), calculate its LFC spectrogram ($\textbf{X}_{LFC}$) with $L$ time steps and $K$ frequency bins;
\item Use MC to predict magnitude of HFC ($\textbf{\^{M}}_{HFC}$) with $N$ frequency bins from the magnitude of $\textbf{X}_{LFC}$, i.e., $\textbf{M}_{LFC}$;
\item Concatenate $\textbf{\^{M}}_{HFC}$ and $\textbf{M}_{LFC}$ as the input to PC for generating the corresponding waveform ($\textbf{\^{x}}_{MelGAN}$);
\item HFC phase ($\textbf{\^{P}}_{HFC}$) can be extracted from $\textbf{\^{x}}_{MelGAN}$, or estimated using FLIP or GLA; 
\item Using $\textbf{\^{P}}_{HFC}$ and $\textbf{\^{M}}_{HFC}$, spectrogram of HFC ($\textbf{\^{X}}_{HFC}$) can be obtained;
\item Concatenate $\textbf{\^{X}}_{HFC}$ and $\textbf{X}_{LFC}$ to generate HR spectrogram ($\textbf{\^{X}}_{HR}$);
\item Finally, HR waveform ($\textbf{\^{x}}_{HR}$) can be obtained from $\textbf{\^{X}}_{HR}$ using iSTFT.
\end{itemize}

The proposed method is compared with the FLIP and GLA methods as different options for HFC phase estimation as presented in Fig.~\ref{fig:proposed-system}.

\subsection{Implementation Details}

For the magnitude components, we adjust the network architecture and configuration in \cite{eskimez2019adversarial} for this task. During the training process, generator of MC takes $\textbf{M}_{LFC}$ with $L=32$ as input. The encoder block contains 4 convolution layers, which have 372,512,512 and 1024 channels with filter size of 7,5,3,3 and stride of 2,2,2,2. The decoder block contains 3 convolution layers, which have 1024, 1024 and 744 channels with filter size of 3,5,7 and stride 1,1,1. To upscale the temporal feature maps in the decoder block, the subpixel layer proposed in Shi et al.~\cite{shi2016real} is used after each LeakyReLU activation. A skip connection is added between each encoder layer and the subpixel output of the corresponding decoder layer. Besides, batch normalization is used after each convolution layer and before the LeakyReLU activation. Following the decoder block, two convolution layers are used to generate the $\textbf{\^{M}}_{HFC}$. They contain $2N$ channels with filter size of 7 and stride 1 and a subpixel layer, $N$ channels with filter size of 9 and stride 1, respectively. As for the discriminator of MC, it consists of three convolution layers, each of which contain 1024 channels with stride of 7, 5 and 3, respectively. These layers are followed by two fully connected layers with 2048 and 1 neurons respectively, and a Sigmoid activation function for the final output of MC.

For the phase component, we use the network architecture described in \cite{kumar2019melgan}. The input of PC is $32 \times 160$ log-mel magnitudes transformed from the MC-predicted linear-scaled magnitudes. PC can then output a waveform of 8192 samples. We only make use of the HFC phase information extracted from the waveform. LFC phase is kept the same as the one calculated from LFC spectrogram of the LR waveform.

\section{Experimental Results}
\label{sec:results}
The dataset we used to train the model consists of 3000 music samples of lossless quality with a sampling rate of 44.1 kHz. These samples can generate the corresponding LR data using low-pass filtering. The spectrum width for LR samples was cut up to 4kHz, while the bandwidth for the target HR was set up to 8kHz. This is known as a 2$x$ task. We used another 25 lossless-LR-paired music samples for evaluating the performance of our proposed system. The STFT is calculated with frame and hop length of 2048 and 256, respectively.

For an objective evaluation between different methods, we choose error metrics of the log-spectral distance (LSD) and the signal-to-noise ratio (SNR). LSD calculates the average spectral distance between the ground-truth and the generated signal in the frequency domain using:
\begin{equation}
  LSD(M,\hat{M}) = \frac{1}{L}\sum_{l=1}^{L}\sqrt{\frac{1}{F}\sum_{f=1}^{F}|M(l,f)-\hat{M}(l,f)|^2},
  \label{eq5}
\end{equation}
where $M(l,f)$ and $\hat{M}(l,f)$ are the ground truth and the estimated log-power spectra at $l$-th time step ($l=1,...,L$) and $f$-th frequency bin ($k=1,...,F$), respectively.



\subsection{The Importance of Phase}

To solely investigate the importance of phase in the audio SR task, we compare FLIP (i.e., phase flipping), GLA, and MelGAN methods with the assumption that we have knowledge of full-band magnitudes and LFC phase. Given the full-band magnitudes ($0-8$kHz) and LFC phase ($0-4$kHz), the task here is to generate phase information of HFC such that HR waveform can be reconstructed. For the FLIP method, the high-frequency phase is produced by flipping the phase of LFC and adding a negative sign. For the GLA method, as introduced in Sec.~\ref{sec:phase_estimation}, a modified version of GLA is used to maintain both the magnitudes and phase of LFC through the iteration process. For the MelGAN method, HFC phase is extracted and complemented by the give magnitudes as well as LFC information in order to obtain the final HR output.


\begin{figure}[!t]
  \centering
  \includegraphics[width=0.9\linewidth]{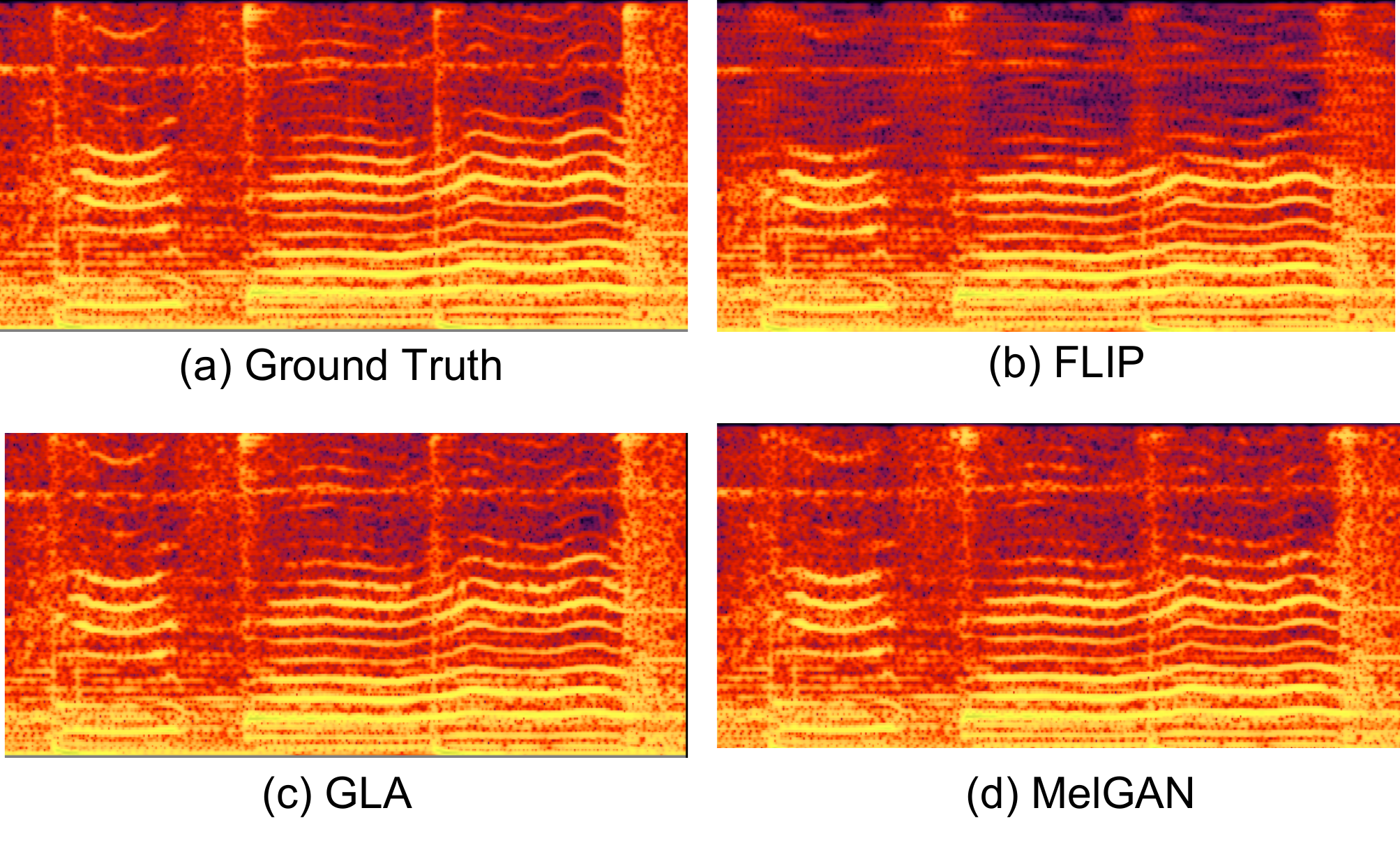}
  \caption{Spectrogram visualization with different phase strategies: Ground truth spectrogram (Top-left), Phase flipping(Top-right), GLA(Bottom-left), and MelGAN(Bottom-right).}
  \label{fig_phase_com}
\end{figure}

\begin{figure}[!t]
  \centering
  \includegraphics[width=0.9\linewidth]{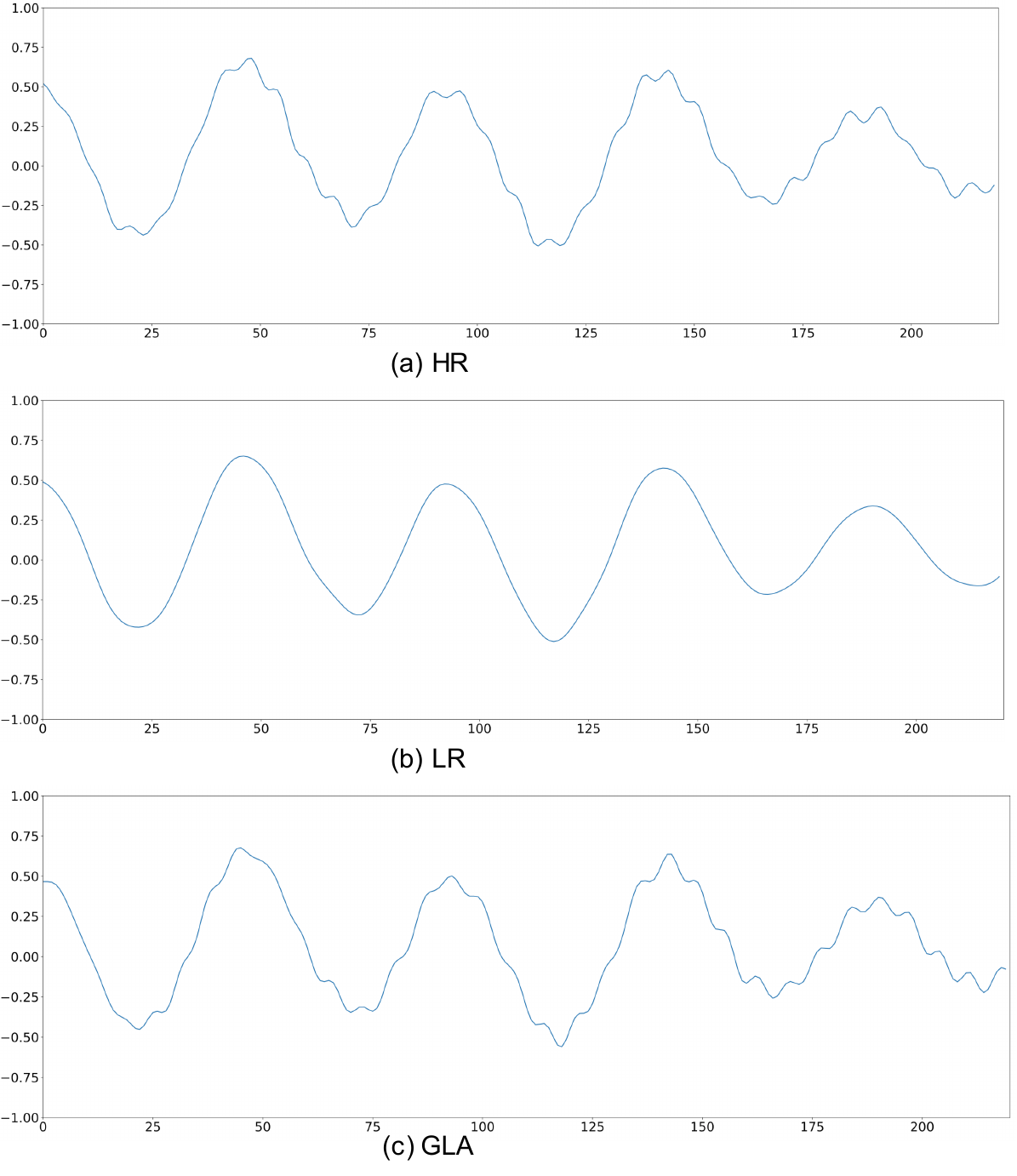}
  \caption{Examples of waveform plotting: Ground truth HR waveform (Top), LR waveform(Middle), Waveform with GLA reconstruction (Bottom)}
  \label{fig_waveform_com}
\end{figure}

\begin{table}[!t]
\centering
\caption{Comparisons between different phase strategies.}
\begin{tabular}{ccccc}
\hline
\textbf{} & \textbf{LSD HF} &\textbf{LSD Full} & \textbf{SNR}\\
\textbf{Method} & \textbf{(dB)} &\textbf{(dB)} &\textbf{(dB)}\\
\hline
LR(input) & $65.78$ &$52.92$& $\bf{18.34}$\\
FLIP\cite{li2015deep,eskimez2019adversarial} & $9.56$ & $7.63$ &$17.95$\\
GLA & $\bf{5.62}$ & $\bf{4.46}$ & $15.39$\\
MelGAN & $6.61$ & $5.27$& $15.86$ \\
\hline
\end{tabular}
  \label{tab_phase_com}
\end{table}
The comparison results of the three phase strategies are presented in Table~\ref{tab_phase_com}. It shows that both the MelGAN and GLA method significantly outperforms the FLIP method used in~\cite{li2015deep,eskimez2019adversarial} in terms of the LSD metric. An example of the generated spectrograms is shown in Fig.~\ref{fig_phase_com} which indicates that FLIP method weakens the energy of HFC due to the poor consistence between magnitudes and the produced phase, while GLA and MelGAN method generate magnitudes close to the ground truth.

It should be noted that the original LR input yields a better SNR value than any other methods used in the comparison. To explain it, we plot an example of ground-truth HR waveform, along with its LR waveform and the reconstructed waveform by GLA as shown in Fig.~\ref{fig_waveform_com}. It presents that bandwidth expansion can effectively generate finer details that are missing in the LR waveform. These details correspond to audible improvements in the high-frequency range. However, the artifacts introduced by the expansion process may severely impact the SNR value. Therefore, the SNR metric may be not suitable for the frequency-domain music SR task. For this reason, we choose the LSD metric alone in the following experiment.

\subsection{Comparison Results of SR Audio}
To evaluate the effectiveness of the proposed joint framework, we implement our approach and compare it with the following state-of-the-art methods:

\begin{itemize}
\item F-DNN: Fully-connected network implementation in the frequency domain (Li et al.~\cite{li2015deep})
\item T-CNN: CNN implementation in the time domain (Kuleshov et al.~\cite{kuleshov2017audio})
\item F-CNN: CNN (generator network of GAN) implementation in the frequency domain (Eskimez et al.~\cite{eskimez2019adversarial})
\item F-GAN: GAN implementation in the frequency domain with phase flipping strategy (Eskimez et al.~\cite{eskimez2019adversarial})
\item Proposed: GAN implementation in the frequency domain for HB magnitude estimation and MelGAN implementation for HB phase estimation.
\end{itemize}


\begin{figure}[!t]
  \centering
  \includegraphics[width=\linewidth]{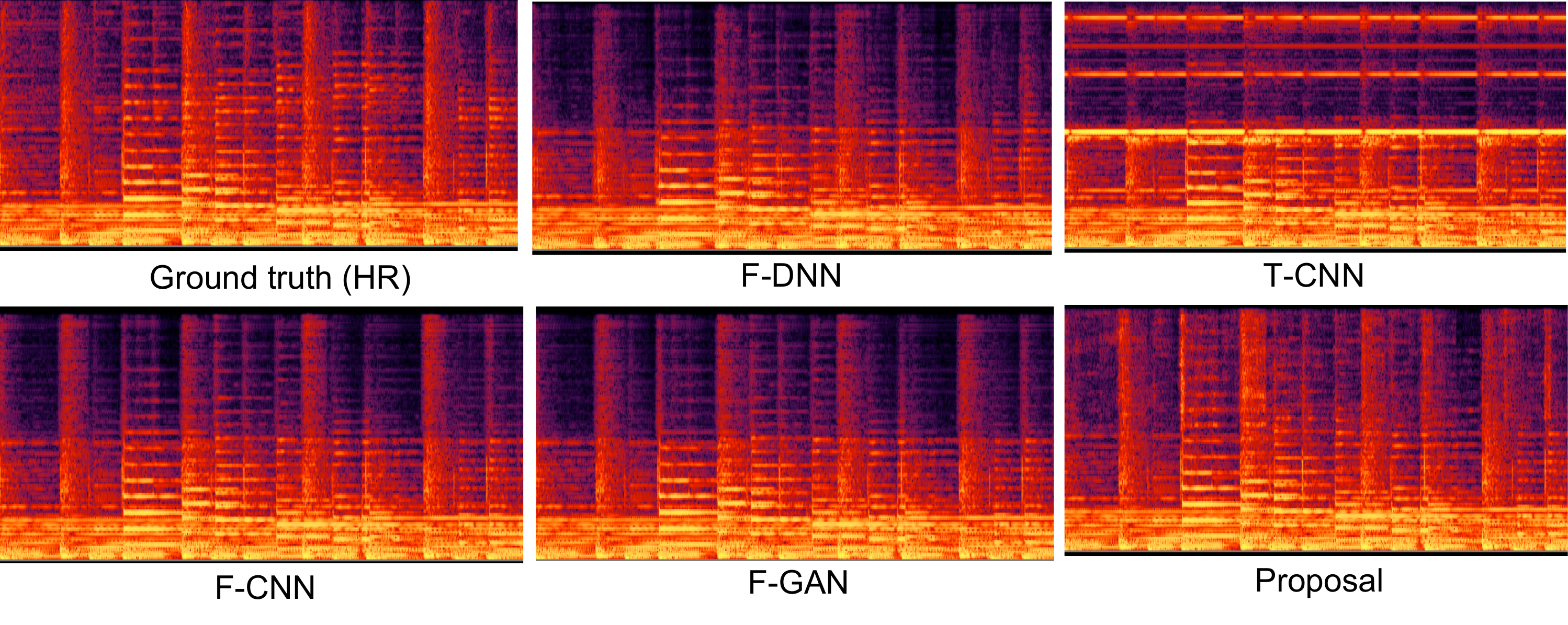}
  \caption{Visualization of predicted spectrograms by our audio SR approach and other existing methods.}
  \label{fig_sr_com}
\end{figure}

\begin{table}[htbp]
\centering
\caption{Comparisons between our audio SR approach and other existing methods.}
\begin{tabular}{cccc}
\hline
\textbf{} & \textbf{LSD HF} &\textbf{LSD Full} \\
\textbf{Method} & \textbf{(dB)} &\textbf{(dB)} \\

\hline
LR(input) & $65.78$ &$52.92$\\
F-DNN\cite{li2015deep} & $11.95$ & $10.03$\\
T-CNN\cite{kuleshov2017audio} & $17.08$ & $14.34$\\
F-CNN\cite{eskimez2019adversarial} & $12.13$ & $9.70$\\
F-GAN\cite{eskimez2019adversarial} & $11.84$ & $9.46$\\
Proposed &$\bf{9.32}$ & $\bf{7.44}$\\
\hline
\end{tabular}
  \label{tab:com_SR}
\end{table}

Fig.~\ref{fig_sr_com} plots the predicted spectrograms for each method. Table \ref{tab:com_SR} presents the comparison results regarding the LSD on high-frequency and full range (denoted as HF and Full). It suggests that our proposed method obviously improves LSD. We achieve the best results among all other existing methods. It should be noted that T-CNN performs much worse than the F-DNN method. This conflicts with the conclusion in \cite{kuleshov2017audio} which conducted the evaluation with speech samples. We believe this is because of the complex nature of polyphonic music, which contains overlapping spectrograms of vocals and various instruments instead of monophonic speech. Thus it is difficult to indirectly expand the frequency bandwidth through temporal interpolation.

We also evaluate the three strategies of phase estimation. This is done by implementing the introduced frequency-domain approaches and replacing FLIP methods with GLA and MelGAN, respectively. In detail, the following methods are evaluated:
\begin{itemize}
\item F-DNN+GLA: Fully-connected network implementation for magnitude prediction(Li et al.\cite{li2015deep}) and GLA phase method for phase estimation
\item F-DNN+MelGAN: Fully-connected network implementation for magnitude prediction (Li et al.\cite{li2015deep}) with MelGAN phase method for phase estimation
\item F-CNN+GLA: CNN implementation\cite{eskimez2019adversarial} for magnitude prediction with GLA method for phase estimation
\item F-CNN+MelGAN: CNN implementation\cite{eskimez2019adversarial} for magnitude prediction with MelGAN method for phase estimation
\item F-GAN+GLA: GAN implementation\cite{eskimez2019adversarial} for magnitude prediction with GLA method for phase estimation
\item F-GAN+MelGAN: GAN implementation\cite{eskimez2019adversarial} for magnitude prediction with MelGAN method for phase estimation.
\end{itemize}

\begin{table}[htbp]
\centering
\caption{Comparisons between audio SR approaches with different phase strategies.}
\begin{tabular}{ccccc}
\hline
\textbf{} & \textbf{LSD HF} &\textbf{LSD Full}\\
\textbf{Method} & \textbf{(dB)} &\textbf{(dB)}\\

\hline
F-DNN+FLIP\cite{li2015deep} & $11.95$ & $10.03$\\
F-DNN+GLA & $10.11$ & $8.63$\\
F-DNN+MelGAN & $9.63$ & $8.26$\\
\hline
F-CNN+FLIP\cite{eskimez2019adversarial} & $12.13$ & $9.70$\\
F-CNN+GLA & $9.67$ & $7.72$\\
F-CNN+MelGAN & $9.58$ & $7.65$\\
\hline
F-GAN+FLIP\cite{eskimez2019adversarial} & $11.84$ & $9.46$\\
F-GAN+GLA &$9.41$& $7.51$\\
F-GAN+MelGAN(proposed) & $\bf{9.32}$ & $\bf{7.44}$\\ 
\hline
\end{tabular}
  \label{tab:com_SR_phase}
\end{table}

The LSD results shown in Table~\ref{tab:com_SR_phase} suggests that the MelGAN method for phase estimation achieves the best result. Both the MelGAN and GLA methods consistently improve the LSD results over the FLIP method used in \cite{li2015deep, eskimez2019adversarial}. Therefore, it is valid that a more accurate phase prediction can effectively solve the music SR tasks.

\section{Conclusions}
In this paper, we investigated the importance of phase in music SR tasks and validated that phase plays an important role in the resulted audio quality. Moreover, we proposed a joint framework which combines a GAN-based mapping method for predicting magnitudes and a MelGAN-based method for predicting the phase information of the high frequency components. Our experimental results show that the proposed approach achieves significantly the best performance comparing with the other state-of-the-art methods in terms of the LSD metric.

\bibliographystyle{IEEEtran}

\bibliography{main}


\end{document}